\documentclass[11pt]{article}
\usepackage{moriond,epsfig,amssymb}

\def\be{\begin{equation}}
\def\ee{\end{equation}}
\def\bea{\begin{eqnarray}}
\def\eea{\end{eqnarray}}

\def\vev#1{\langle #1 \rangle}
\begin{document}
\vspace*{4cm}
\title{CP from strings: ideas and problems}

\author{Thomas Dent}

\address{University of Michigan, Randall Lab., 500 E.\ University Ave.,\\
Ann Arbor, MI 48109 U.\,S.\,A.}

\maketitle\abstracts{\noindent CP has a natural embedding in 
superstring models as a gauge symmetry involving inversion of the 
compactified space. Hence the source of CP violation could be geometrical. 
Such models face the problem of how to suppress contributions to fermion 
electric dipole moments from softly-broken supersymmetry, as well as 
other problems of string phenomenology. Stringy symmetries are useful 
in evaluating models, and rule out some scenarios.}

\noindent
In this talk I will describe some aspects of CP violation within string
models and low-energy supersymmetry (SUSY): the two are connected since
it is extremely difficult to construct consistent string models without 
some supersymmetry. First I describe some established results which show
that CP is an exact symmetry of the underlying theory, and has an elegant
interpretation in geometrical terms. This motivates models of spontaneous
CP violation through geometric moduli, which appear in the low-energy 
effective field theory as scalar singlets with nonrenormalisable couplings
to Standard Model (SM) fields. I describe the ``SUSY CP problem'', the
fact that superpartner loops are expected to generate fermion electric 
dipole moments (EDM's) in excess of experimental bounds, and discuss some
possible solutions. I also remind the reader of the problems encountered
by any attempt to describe details of low-energy physics within string 
theory. Finally I report a recent result which allows string model 
results to be interpreted without detailed calculation, and in many cases
ruled out as sources of CP violation.

\section{CP as gauge symmetry and as geometry}
CP symmetry in four dimensions reverses the chirality and U$(1)$
charges of fermions and the orientation of spacetime: surprisingly, if 
we consider the ten-dimensional spacetime and enlarged gauge group of 
heterotic or Type I string theory, CP can be embedded as a discrete 
subgroup of the gauge symmetry of the full theory \cite{CPgauge}. 
Fermion representations in the 10d
Lorentz group must be Majorana, since the fermion must live in the same 
irrep as its complex conjugate: in fact, $\mathcal{N}=1$ string has 
Majorana-Weyl fermions. Gauge group representations must 
be such that each irrep containing a field with U$(1)$ charge(s) $q_a$ 
also contains a field with charge $-q_a$: the irreps of $E_8$ and 
SO$(32)$, among others, satisfy this condition automatically. Finally, 
the 4d parity transformation must correspond to a proper Lorentz 
transformation in 10d, which is achieved by taking CP to act by 
inversion on the 6-dimensional compact space, by a 6-dimensional Lorentz
transformation with negative determinant \footnote{It was already noted 
in \cite{StroWitten} that this inversion was a natural realization of CP.}. 
As a discrete gauge symmetry, CP is conserved even nonperturbatively in
string interactions, hence CP-violating couplings including 
$\theta_{\rm QCD}$ are in principle calculable given a particular mechanism
for symmetry-breaking at low energies.

An intriguing possibility is for CP to be broken by extra-dimensional 
geometry: if there is no inversion under which the 6-dimensional space, 
possibly including vector or tensor field backgrounds along the compact 
directions, is invariant, CP is violated spontaneously. One would expect 
low-energy couplings which depend on this geometry, for example quark 
Yukawas or soft SUSY-breaking terms, to be complex. In general, not enough
is known about the 6d space for the low-energy parameters to be found 
explicitly (even the metric of Calabi-Yau spaces is in general unknown), 
but in simple, somewhat unrealistic examples this can be done, with 
interesting results which may have implications for models closer to the
real world.

Orbifolds \cite{BL} are the most intensively-studied examples of 
string compactification with $\mathcal{N}=1$ supersymmetry: they are simply 
$2n$-dimensional flat space with identifications under the action of a
discrete group. For abelian orbifolds this group consists of six translation 
generators, which convert $\mathbb{R}^6\simeq \mathbb{C}^3$ into the 
torus $T^6$, and a (product of) discrete rotations $Z_N(\times Z_M)$
which acts as $z^\alpha\mapsto e^{2\pi i k^\alpha/ N(M)}z^\alpha$ on the 
three complex coordinates $z^\alpha$, $\alpha=1,2,3$. Then CP acts as 
$z^\alpha\mapsto z^{\alpha*}$ (which may be in general be combined with 
a 6-dimensional rotation). In any orbifold the antisymmetric tensor field 
$B^{MN}$ in the compact directions is a free parameter which is easily
seen to be CP-odd. For complex planes where the orbifold rotation is $Z_2$, 
the relative orientation of the two basis vectors of the torus (defined as
$1$ and $\tau^\alpha$ when normalised to the overall radius $R$) can be 
CP-odd if Re$\,\tau^\alpha\neq 0$. By convention, these quantities are 
combined with other geometric parameters into the K{\" a}hler 
moduli $T^\alpha= 2(R^{\alpha2}+iB^\alpha)$ where $B^\alpha=B^{MN}$, 
$M=N-1=2\alpha-1$, and the complex structure moduli $U^\alpha\sim 
\tau^\alpha/i$, which transform into their complex conjugates 
under CP. Then in simple orbifold models, Yukawas and soft terms are 
relatively easy to compute as functions of $T^\alpha$ and $U^\alpha$.
Unfortunately, none of these models gives a good account of the quark
sector without further elaboration: either there are more than three 
generations, or the Yukawa couplings allowed by selection rules cannot
give rise to a CKM phase (see {\em e.g.}\/\ \cite{Lebedev01}). Then either
the extra generations must be decoupled in some way, or higher-order 
nonrenormalizable couplings invoving charged scalar v.e.v.'s must play
a role in generating the Yukawa textures ({\em e.g.}\/\ \cite{Giedt02}): 
either alternative is likely to make the models more difficult, if not 
impossible, to compute reliably.

\subsection{CP as scalar v.e.v.'s}
Supersymmetric string vacua have vanishing cosmological constant to all
orders in perturbation theory, therefore the effective potential on the
space of continuous parameters, including the moduli, is flat. Thus one
can consider $T^\alpha$ and $U^\alpha$ as scalar fields massless before
SUSY-breaking, which appear in the low-energy effective theory 
\footnote{The possibility also exists \cite{me_modinv,me_nuclphys} that 
CP is broken by {\em discrete}\/ parameters of the compactification, which 
would not correspond to scalar degrees of freedom.}.
After supersymmetry-breaking in a hidden sector by some nonperturbative 
mechanism such as gaugino condensation, the moduli (and in general the 
dilaton $S$, which gives the gauge kinetic function at tree level) become
stabilized in a nontrivial potential and receive $F$-terms, resulting 
in soft breaking in the observable sector.
The moduli couple in general in a flavour-dependent way to MSSM matter,
thus the expectation is for nonuniversality and CP violation in the soft
terms (even if soft scalar masses are degenerate), which would 
likely lead to new physics signals beyond the ``minimal flavour 
violation'' benchmark. The apparent lack of such signals in CP and 
flavour-changing observables thus represents a challenge for such models.

\section{Problems of CP violation in SUSY}
This challenge is most marked for fermion EDM's, where extremely sensitive
experimental limits exist \cite{EDMexp}: the SM contribution is negligible, 
but superpartner loops induce EDM's two or more orders of magnitude
above the limits, for soft terms of a few hundred GeV in magnitude with 
complex phases of order unity \cite{SUSYEDM}. The soft terms involved are 
the gaugino masses $M_{i}\lambda_i\lambda_i$, $i=1,2,3$, the scalar 
bilinear coupling $B\mu H_U H_D$ (where $\mu H_U H_D$ is the effective 
superpotential coupling generating Higgsino masses) and the scalar
trilinear $A$-terms $\hat{A}^{u}_{ij} \tilde{q}_i\tilde{u}^c_j H_U
+ (u\!\rightarrow\!d) + (q\!\rightarrow\! l)$ where the down quark and 
charged lepton terms have a similar form. The combinations $BM^*_i|\mu|/|B|$ 
and $\hat{A}^{\rm SCKM}_{11}M^*_i$ are invariant under phase redefinitions 
and appear directly in the expressions for loop-induced EDM's. Here 
$\hat{A}^{\rm SCKM}_{11}$ is the (11) element, for the up, down and lepton
sectors, in the ``super-CKM'' basis where fermion masses and 
fermion-sfermion-gaugino couplings are diagonal. Then 
the phase of $BM^*_i$ should be less than $10^{-2}$, and the imaginary parts
of $\hat{A}^{\rm SCKM}_{11}M^*_i$ somewhat less than $10^{-6}$, to satisfy
experimental bounds, in the absence of cancellations which appear 
increasingly unlikely given improved bounds on the mercury EDM 
(see \cite{SUSYEDM}, third reference). In the limit
of universal soft terms, the problem remains but is less severe for the 
$A$-terms: the (11) element is now proportional to the small quark or
lepton masses $m_{u,d,e}$, automatically inducing a suppression of order 
$10^{-5}$. Still, in almost all cases the soft phases have to be small, 
in contrast to the large Yukawa phases which are usually taken to generate 
CP violation in the quark sector. Radiative effects calculated by RG 
running can have an important effect on these bounds, since the $A$-terms 
obtain a large contribution proportional to gaugino masses, which 
``dilutes'' the imaginary parts of $\hat{A}M^*_i$, on running from the GUT 
to the electroweak scale; the large ``top'' and ``bottom'' entries 
$\hat{A}^{u,d}_{33}$ also give contributions to the imaginary parts of 
$\hat{A}_{11}$ and $B\mu$, which may be significant.

There are many ways to get around this problem in SUSY, but none 
of them is strictly motivated within string theory. Perhaps the most
obvious is to arrange for the scalar superpartners to be 
heavy \cite{heavies} (a few TeV is sufficient) suppressing the loop 
diagrams; this can be done without unacceptable fine-tuning of the
electroweak sector if the third generation superpartners remain light 
\footnote{However, new contributions to EDM's have recently been 
found \cite{newEDM} that do not decouple in the heavy superpartner limit 
and may be dangerous for order (1) phases.}.

If we allow an extended gauge group, one simple solution is to embed 
MSSM into a parity-symmetric (left-right) SUSY model at the seesaw scale 
\cite{Mohapatra_etal}. 
This model leads to hermitian Yukawas and $A$-terms, real gluino masses, 
$\mu$ and $B\mu$ terms so that it can solve both the SUSY CP and strong 
CP problems; many authors have pointed out that left-right symmetric 
models can emerge from strings, although it is not clear that left-right
symmetry can be broken in the required way in these constructions.

``Approximate CP'' is the proposal, motivated by spontaneous breaking of exact
CP symmetry, that all complex phases are small: suppressed EDM's are thus 
natural in the t'Hooft sense, but the KM phase is also taken to be small, 
thus one requires additional sources of CP violation in flavour-changing 
interactions \cite{approximate}. This is possible, if somewhat complicated,
to arrange in the $K^0$ system, consistent with small phases, but the 
prediction for the $B_d$ decay asymmetry $a_{J/\Psi K_S}$ is small ($<0.1$), 
thus approximate CP as usually formulated is ruled out by recent results 
\cite{B's}. However, it is worth noting that the effect of small Yukawa 
phases is strongly dependent on the model of flavour ({\em i.e.}\/\ on the 
basis of quark states in which the phases are introduced). It is even
possible to formulate a large KM phase as a small perturbation away from
a CP-invariant theory, since it can result from Yukawa phases of order 
$10^{-3}$ in a democratic theory of flavour (where all Yukawa couplings
are approximately equal): the Jarlskog parameter $J\simeq 2\times 
10^{-5}$ is then a more appropriate measure of CP violation in the SM.
One can also reformulate approximate CP by imposing small imaginary 
parts, which would generically result in small soft phases and a large 
KM phase.

\section{Problems of strings}
We already noted that it is difficult to generate correct Yukawa textures 
reliably in three-generation heterotic models. ``Pseudo-anomalous'' 
U(1) groups, where the anomaly is cancelled by a Green-Schwarz mechanism, 
are common (see {\em e.g.}\/\ \cite{Giedt02}): in order to cancel the 
resulting F-I D-term, some charged scalars get 
v.e.v.'s, thus in the stabilized vacuum, nonrenormalizable superpotential 
terms may contribute significantly to the effective Yukawa couplings. The 
form of such terms is prohibitively hard to calculate, so one can do little
more than estimate their magnitude: note that charged scalar v.e.v.'s
may also be a source of CP violation. In the presence of many potential
sources, a reasonable strategy is to evaluate each one in turn and only
consider more complicated cases when the simple ones are ruled out. It is
in this spirit that recent investigations of moduli as the source of CP
violation have been carried out.

More serious and widereaching is the problem of vacuum selection, which
encompasses both the choice between discrete (topologically different) 
compactifications and the stabilization of the many continuous flat 
directions. Without a model in which the SM or MSSM is obtained ``cleanly'' 
({\em i.e.}\/\ without extra or exotic matter), one is usually reduced 
to calculating Yukawas and soft terms with a somewhat arbitrary 
assignment of the (MS)SM fields, thus apart from very general properties 
it is unclear how much credence should be given to the results.

Supersymmetry-breaking is another, related area of uncertainty \cite{BL}: 
since it is necessarily nonperturbative in nature, it cannot be treated 
directly in string theory, thus one must use low-energy approximations 
which may hide some important features. 
The constraints that must be satisfied by any mechanism are to stabilize
the moduli, dilaton and other scalars, with sufficiently large masses, 
and give soft masses to the superpartners consistent with bounds from 
electroweak symmetry-breaking, FCNC, {\em etc.}; in addition, one requires 
to the vacuum energy to vanish to good accuracy for the model to be 
self-consistent. To do all this turns out to be extremely 
difficult, such that one needs to introduce additional parameters, whose
meaning in terms of string theory is unclear, and which must be set by 
hand to get reasonable results.

\section{Results from stringy symmetry}
\begin{figure}
\hspace{3cm}
\psfig{figure=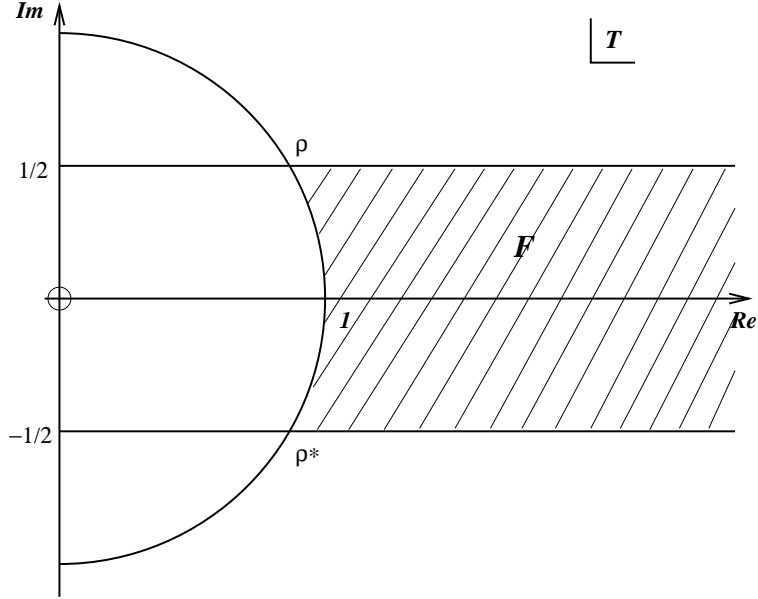,height=8cm}
\caption{The fundamental domain of SL$(2,\mathbb{Z})$ acting on $T$
\label{fig:F}}
\end{figure}

Still, one can use the special properties of string models to diagnose 
whether some specific scenarios of CP violation are viable, without 
detailed calculation.
Many simple heterotic compactifications have a duality group $\Gamma$ 
acting on the geometric moduli: this means that doing string perturbation 
theory on one background $\{T,U\}$ gives the same result as on another 
background $\{\gamma(T),\gamma(U)\}$, where $\gamma(T(U))$ is the action
of a group element $\gamma$ on $T(U)$ (suppressing the $\alpha$ labels).
If we restrict attention to group elements which do not mix complex planes,
then we always have three separate SL$(2,\mathbb{Z})$ groups acting on
the $T^\alpha$ as
\be
\gamma(a^\alpha,b^\alpha,c^\alpha,d^\alpha):\ T^\alpha \mapsto \frac{a^\alpha T^\alpha - i b^\alpha}{i c^\alpha T^\alpha + d^\alpha},\ a^\alpha d^\alpha-  b^\alpha c^\alpha=1
\ee
for integer $a,b,c,d$, generated by $T\mapsto 1/T$ and $T\mapsto T+i$. 
Then, the low-energy physics at one v.e.v.\ $\vev{T}=T_0$ should
be the same as at $\vev{T}=\gamma(T_0)$. Calculations of $T$-dependent 
gauge and Yukawa couplings are consistent with this symmetry, called
{\em modular invariance}. The low-energy supergravity effective is 
invariant under this replacement as expected, when matter fields are also
given appropriate transformation properties. All possible values of $T$,
and therefore all possible values of Yukawas and soft terms for a
particular string model, can be reached by a modular transformation 
from precisely one point inside the fundamental domain $\mathcal{F}$, 
depicted in Fig.~\ref{fig:F}. 

Then, given the exact CP symmetry of the underlying theory, it is relatively
simple to show that moduli v.e.v.'s on the boundary of $\mathcal{F}$ cannot
be a source of CP violation \cite{me_modinv}, since they are modular 
images of their own complex conjugates (under $T\mapsto 1/T$ or 
$T\mapsto T+i$), therefore describe backgrounds for which physics is 
unchanged by CP transformation. If SM matter fields are in twisted
sectors, which are localised at fixed points of the $Z_N(\times Z_M)$ group,
then they mix with each other under duality, ensuring that the total 
effective Lagrangian is invariant: this mixing, which is unavoidably 
nonabelian \cite{Lauer:1990tm}, makes it somewhat subtle to prove the result, 
particularly in the case of models with more than three generations where 
some twisted matter decouples. 

If one assumes that the extra states become heavy in the original 
``twist'' basis without mixing with the three light generations, 
then the CKM matrix might not in principle be a modular invariant smooth 
function of $T$, and CP-violating quantities might not vanish on 
$\mathcal{F}$, because the modular transformation would not act unitarily 
on the light generations \cite{Lebedev01}. Then one would have to 
switch to a different basis of light states after modular transformation, 
for which the CKM matrix was a different function of $T$; but the value of 
the new function at $\vev{T}=\gamma(T_0)$ would be equal to the value of 
the old function for $\vev{T}=T_0$, confirming that physics is unchanged 
under duality. In fact, in any reasonable situation, mass matrices in the 
twist basis must have off-diagonal entries, and thus are diagonalised by 
{\em modulus-dependent}\/ matrices. As a result, the mass eigenstates are 
formally {\em invariant}\/ under duality: thus, the couplings of
the Lagrangian written in the mass eigenstate basis are smooth 
SL$(2,\mathbb{Z})$ invariant functions of $T$, and the CP-odd part of any 
such function vanishes on the boundary of $\mathcal{F}$. 
The dilaton may be a source
of CP violation consistent with modular symmetry, but it cannot play
a significant role in the observed effects, since its couplings cannot 
induce CP violation in flavour-changing terms \cite{me_modinv}. The 
construction of modular invariant eigenstates may have interesting
consequences for flavour in models with twisted matter.

This result is significant because most models of modulus stabilization 
consistent with modular invariance result either in real $\vev{T}$ or 
in exactly these boundary values, 
(with, however, some 
exceptions \cite{Bailin+KhalilLM,me_nuclphys}). It has been suggested 
that complex values of $T$ on the unit circle could form part of a 
solution of the SUSY CP problem, since $F^T$ then vanishes in most 
cases, but then there must be additional sources of CP violation in the 
model, which may or may not give additional contributions to soft terms.
If $\vev{T}$ is sufficiently close to the boundary, then CP 
violation in both Yukawas and soft terms is expected to be small, motivating
approximate CP; however, as remarked above, this scenario is likely ruled 
out without some special type of flavour structure.


\section*{Acknowledgments}
This research was supported in part by DOE Grant DE-FG02-95ER40899 Task G.



\section*{References}
\begin{thebibliography}{99}

\bibitem{CPgauge}
M.~Dine, R.\,G.~Leigh and D.\,A.~MacIntire,
Phys.\ Rev.\ Lett.\ {\bf 69} (1992) 2030;
K-W.~Choi, D.\,B.~Kaplan and A.\,E.~Nelson,
Nucl.\ Phys.\ B {\bf 391} (1993) 515.

\bibitem{StroWitten} A.~Strominger and E.~Witten,
Commun.\ Math.\ Phys.\ {\bf 101} (1985) 341.

\bibitem{BL} For a review, see D.~Bailin and A.~Love,
Phys.\ Rept.\ {\bf 315} (1999) 285.

\bibitem{Lebedev01} O.~Lebedev,
Phys.\ Lett.\ B {\bf 521} (2001) 71 [hep-th/0108218].

\bibitem{Giedt02} J.~Giedt,
hep-ph/0204017,
Annals Phys.\ {\bf 297} (2002) 67 [hep-th/0108244].

\bibitem{me_modinv} T.~Dent,
Phys.\ Rev.\ D {\bf 64} (2001) 056005 [hep-ph/0105285];
JHEP {\bf 0112} (2001) 028 [hep-th/0111024].

\bibitem{me_nuclphys} T.~Dent,
Nucl.\ Phys.\ B {\bf 623} (2002) 73 [Erratum-ibid.\ B {\bf 629} (2002) 493]
[hep-th/0110110].

\bibitem{EDMexp} P.\,G.~Harris {\it et al.},
Phys.\ Rev.\ Lett.\ {\bf 82} (1999) 904;
E.\,D.~Commins, S.\,B.~Ross, D.~DeMille and B.\,C.~Regan,
Phys.\ Rev.\ A {\bf 50} (1994) 2960;
M.\,V.~Romalis, W.\,C.~Griffith and E.\,N.~Fortson,
Phys.\ Rev.\ Lett.\  {\bf 86} (2001) 2505 [hep-ex/0012001].

\bibitem{SUSYEDM} 
S.~Abel, S.~Khalil and O.~Lebedev,
Nucl.\ Phys.\ B {\bf 606} (2001) 151 [hep-ph/0103320];
T.~Falk, K.\,A.~Olive, M.~Pospelov and R.~Roiban,
Nucl.\ Phys.\ B {\bf 560} (1999) 3 [hep-ph/9904393];
V.~D.~Barger {\em et al.}, 
Phys.\ Rev.\ D {\bf 64} (2001) 056007 [hep-ph/0101106];
S.~Pokorski, J.~Rosiek and C.~A.~Savoy,
Nucl.\ Phys.\ B {\bf 570} (2000) 81 [hep-ph/9906206].

\bibitem{heavies} Y.~Kizukuri and N.~Oshimo,
Phys.\ Rev.\ D {\bf 45} (1992) 1806,
Phys.\ Rev.\ D {\bf 46} (1992) 3025;
A.\,G.~Cohen, D.\,B.~Kaplan and A.\,E.~Nelson,
Phys.\ Lett.\ B {\bf 388} (1996) 588 [hep-ph/9607394].

\bibitem{newEDM}
D.~Chang, W.\,F.~Chang and W.\,Y.~Keung,
hep-ph/0205084;
O.~Lebedev and M.~Pospelov,
hep-ph/0204359.

\bibitem{Mohapatra_etal} R.~N.~Mohapatra and A.~Rasin,
Phys.\ Rev.\ D {\bf 54} (1996) 5835 [hep-ph/9604445];
K.~S.~Babu, B.~Dutta and R.~N.~Mohapatra,
Phys.\ Rev.\ D {\bf 65} (2002) 016005 [hep-ph/0107100].

\bibitem{approximate}
S.\,A.~Abel and J.~M.~Fr{\` e}re,
Phys.\ Rev.\ D {\bf 55} (1997) 1623 [hep-ph/9608251].
G.~Eyal and Y.~Nir,
Nucl.\ Phys.\ B {\bf 528} (1998) 21 [hep-ph/9801411].
M.~Dine, E.~Kramer, Y.~Nir and Y.~Shadmi,
Phys.\ Rev.\ D {\bf 63} (2001) 116005 [hep-ph/0101092].

\bibitem{B's} B.~Aubert {\it et al.} [BABAR Collaboration],
Phys.\ Rev.\ Lett.\ {\bf 87} (2001) 091801 [hep-ex/0107013].
K.~Abe {\it et al.} [Belle Collaboration],
Phys.\ Rev.\ Lett.\ {\bf 87} (2001) 091802 [hep-ex/0107061].

\bibitem{Lauer:1990tm}
J.~Lauer, J.~Mas and H.~P.~Nilles,
Nucl.\ Phys.\ B {\bf 351} (1991) 353.

\bibitem{Bailin+KhalilLM}
D.~Bailin, G.\,V.~Kraniotis and A.~Love,
Nucl.\ Phys.\ B {\bf 518} (1998) 92 [hep-th/9707105],
Phys.\ Lett.\ B {\bf 483} (2000) 425 [hep-th/0004052];
S.~Khalil, O.~Lebedev and S.~Morris,
hep-th/0110063.

\end{thebibliography}
\end{document}